\documentclass[twocolumn,aps,prb,superscriptaddress,longbibliography]{revtex4-1}
\usepackage[colorlinks,bookmarks=false,citecolor=blue,linkcolor=red,urlcolor=blue]{hyperref}
\usepackage{verbatim}
\usepackage{color}
\usepackage{amsmath,amssymb,bm,braket,float,mathtools}
\usepackage{graphicx,xfrac,appendix}
\usepackage[english]{babel}
\usepackage{epstopdf}
\newcommand{\mycite}[1]{\scalebox{1.3}[1.3]{\raisebox{-0.65ex}{\cite{#1}}}}

\begin{document}

\title{Thermodynamic modes of a quasiperiodic mobility-edge system in a quantum Otto cycle}

\author{Ao Zhou}
\affiliation{Department of Physics, Zhejiang Normal University, Jinhua 321004, China}
	
\author{Shujie Cheng}
\thanks{chengsj@zjnu.edu.cn}
\affiliation{Xingzhi College, Zhejiang Normal University, Lanxi 321100, China}
\affiliation{Department of Physics, Zhejiang Normal University, Jinhua 321004, China}
	
\author{Gao Xianlong}
\thanks{gaoxl@zjnu.edu.cn}
\affiliation{Department of Physics, Zhejiang Normal University, Jinhua 321004, China}

\date{\today}

\begin{abstract}
We investigate thermodynamic operation of a quasiperiodic lattice with an exact mobility edge, described by the Biddle--Das Sarma model.
We use this model as the working medium of a quantum Otto cycle and map its operating mode as a function of the hopping-range parameter $p$,
the initial and final potential strengths $V_i$ and $V_f$, and two idealized protocols for the isolated strokes.
In a near-adiabatic (state-frozen) protocol, where the density matrix is approximately unchanged during the isolated strokes, the cycle supports only
two modes: a \emph{heater} and an \emph{accelerator}. In an adiabatic protocol, where level populations are preserved while the spectrum is deformed,
two additional modes appear: a \emph{heat engine} and a \emph{refrigerator}. Our results show that mobility-edge systems can realize multiple
thermodynamic functions within a single platform and provide guidance for switching between modes by tuning $p$, $V_i$, and $V_f$.
\end{abstract}

\maketitle

\section{Introduction}

Quantum thermal devices were first placed on a clear footing by the proposal of a quantum heat engine based on a three-level maser \cite{scovilThreeLevelMasersHeat1959}.
Since then, quantum heat engines and refrigerators have been studied extensively, both as models of energy conversion at small scales and as testbeds for nonequilibrium quantum thermodynamics
\cite{kieuQuantumHeatEngines2006, heThermalEntangledQuantum2012a, quanQuantumThermodynamicCycles2007, zhangFourlevelEntangledQuantum2007, scullyExtractingWorkSingle2003, zhangEntangledQuantumHeat2008, henrichDrivenSpinSystems2007, huangSpecialEntangledQuantum2013, wangThermalEntangledQuantum2012, wangThermalEntanglementTwoatom2009, solfanelliQuantumHeatEngine2023g, shankarNearestneighborFrustratedRandombond1987, thomasCoupledQuantumOtto2011, altintasGeneralFormalismLocal2015, feldmannQuantumFourstrokeHeat2003, ivanchenkoQuantumOttoCycle2015, azimiQuantumOttoHeat2014, hubnerSpindependentOttoQuantum2014, kollasExactlySolvableRelativistic2024c, chenBoostingPerformanceQuantum2019c, cakmakLipkinMeshkovGlickModelQuantum2016c, leggioOttoEngineIts2016, tuncerQuantumIsingSpinGlass2024c, wangQuantumOttoEngine2012c, piccittoIsingCriticalQuantum2022c, rodinThreedimensionalHarmonicOscillator2024c, mitchisonQuantumThermalAbsorption2019}.
Many studies adapt familiar cycles---Carnot \cite{quanQuantumclassicalTransitionPhotonCarnot2006}, Otto \cite{xuUniversalQuantumOtto2024d, chandFinitetimePerformanceSingleion2021c},
Brayton \cite{wuGeneralizedModelOptimum2006, huangQuantumBraytonCycle2013}, and Diesel \cite{PhysRevB.88.214421}--by replacing the classical working substance with a controllable quantum system.
The cycle then converts heat and work according to quantum dynamics and thermalization \cite{kosloffQuantumThermodynamicsDynamical2013}.

A recurring theme is that the operating mode and performance can depend strongly on properties that have no classical counterpart, such as coherence \cite{scullyExtractingWorkSingle2003},
dissipation mechanisms \cite{hardalSuperradiantQuantumHeat2015, quanQuantumclassicalTransitionPhotonCarnot2006}, interactions \cite{zhang_entangled_2008},
and quantum critical points \cite{tuncerQuantumIsingSpinGlass2024c}. This sensitivity suggests that systems with sharp spectral reorganizations may offer useful control knobs for switching between
thermodynamic functions.

Quasiperiodic lattices provide a concrete setting where spectral and eigenstate properties can be tuned in a controlled way, and they have been explored as working media for quantum thermal cycles
\cite{chiaracaneQuasiperiodicQuantumHeat2020a, xuQuantumTransportProperty2025, suoWignerDistributionWigner2026}. In particular, recent work indicates that changes in eigenstate character
(extended versus nonextended) can be correlated with changes in thermodynamic operating mode \cite{xuQuantumTransportProperty2025}. For quasiperiodic models without mobility edges, a single
Hamiltonian can support four modes (heat engine, refrigerator, heater, and accelerator) by tuning system parameters \cite{suoWignerDistributionWigner2026}. By contrast, for quasiperiodic
\emph{mobility-edge} systems--where extended and localized states coexist at different energies--the thermodynamic landscape is less explored, and existing work has focused mainly on heat-engine operation \cite{chiaracaneQuasiperiodicQuantumHeat2020a}. This motivates two practical questions: Which thermodynamic modes are accessible in a mobility-edge system, and under what conditions do they emerge?

In this paper we address these questions using the Biddle--Das Sarma model as a representative quasiperiodic mobility-edge system \cite{BS_model}. We embed the model in a quantum Otto cycle and
construct mode diagrams as functions of the hopping-range parameter $p$ and the potentials $V_i$ and $V_f$, comparing a near-adiabatic (state-frozen) protocol and an adiabatic protocol for the isolated strokes.
Section~\ref{S2} summarizes the model and the mobility-edge condition, together with the localization measures used.
Section~\ref{S3} defines the Otto cycle, the heat and work conventions, and the two protocols for the isolated strokes.
Section~\ref{S4} presents the resulting mode diagrams and discusses how mode competition depends on $p$, $V_i$, and $V_f$.
Section~\ref{S5} concludes with a summary and outlook.

\section{Biddle--Das Sarma model}\label{S2}

We consider the one-dimensional Biddle--Das Sarma model, introduced as a minimal quasiperiodic lattice with an exact mobility edge under exponentially decaying hopping \cite{BS_model}.
The Hamiltonian reads
\begin{equation}
\hat{H}=\sum_{n,n'\neq n} t\,e^{-p|n-n'|}
\left(\hat{c}^{\dag}_{n'}\hat{c}_{n}+\hat{c}^{\dag}_{n}\hat{c}_{n'}\right)
+\sum_{n} V_{n}\,\hat{c}^{\dag}_{n}\hat{c}_{n},
\label{eq5}
\end{equation}
where $\hat{c}_n^\dagger$ ($\hat{c}_n$) creates (annihilates) a particle at site $n$, $t$ sets the energy scale, and $p>0$ controls how rapidly the hopping decays with distance.
The onsite potential is quasiperiodic,
\begin{equation}
V_{n}=V\cos(2\pi\alpha n),
\end{equation}
with irrational $\alpha$ (we use $\alpha=(\sqrt{5}-1)/2$).

A defining property of the model is an analytic mobility edge obtained from self-duality \cite{BS_model},
\begin{equation}
\frac{E_{c}}{t}=V\cosh(p)-1,
\end{equation}
which separates extended from localized single-particle eigenstates at energy $E_c$.

To characterize eigenstate localization numerically, we use the inverse participation ratio (IPR) \cite{BS_model}.
For a normalized eigenstate $\ket{\psi^{(j)}}=\sum_{n=1}^{L}\phi^{(j)}_{n}\,\hat{c}^{\dag}_{n}\ket{0}$, the IPR is
\begin{equation}
{\rm IPR}^{(j)}=\sum_{n=1}^{L}\left|\phi^{(j)}_{n}\right|^4.
\end{equation}
Extended states satisfy ${\rm IPR}\sim 1/L$, while localized states yield ${\rm IPR}=O(1)$ as $L$ increases.
We also report the associated fractal dimension
\begin{equation}
D^{(j)}=-\frac{\log {\rm IPR}^{(j)}}{\log L},
\end{equation}
which approaches $1$ for extended states and $0$ for localized states in the large-$L$ limit.

\begin{figure}[htp]
\includegraphics[width=0.5\textwidth]{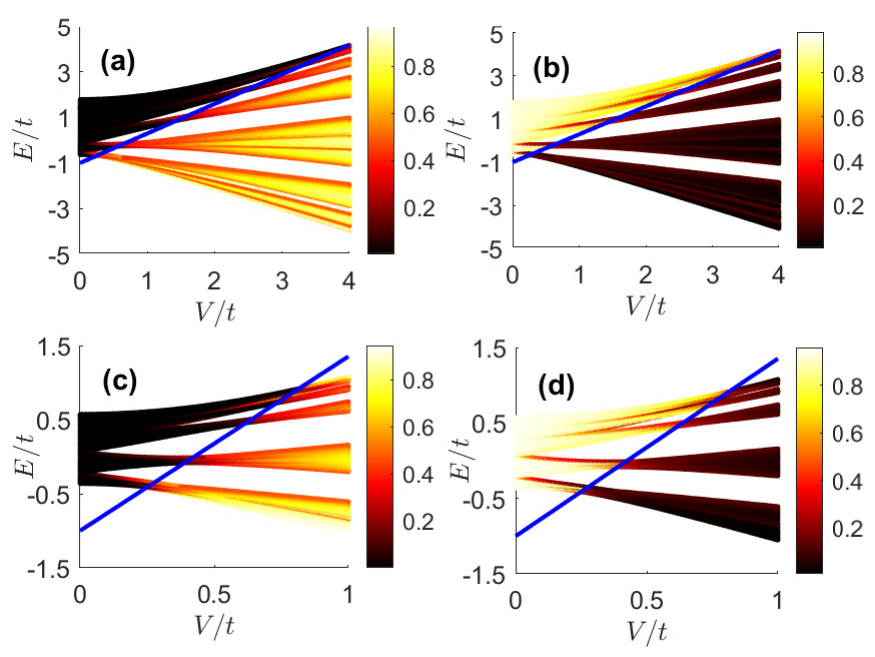}\\
\caption{(Color online) (a) ${\rm IPR}$ at each energy level versus $V/t$ for $p=0.75$ and $L=610$.
(b) Fractal dimension $D$ versus $V/t$ for the same parameters.
(c) ${\rm IPR}$ versus $V/t$ for $p=1.5$ and $L=610$.
(d) Fractal dimension $D$ versus $V/t$ for the same parameters. 
The color in (a) and (c) represent the ${\rm IPR}$ and that in (b) and (d) denote the fractal dimension $D$. 
The solid blue line denotes the mobility edge $E_{c}/t = V \cosh(p)-1$.}
\label{f1}
\end{figure}

Figure~\ref{f1} shows ${\rm IPR}$ and $D$ across the spectrum as functions of $V/t$ for two representative values of $p$.
The mobility-edge line cleanly separates regions dominated by extended or localized eigenstates.
A practical implication for the Otto-cycle setting is that changing $p$ shifts the mobility edge: for smaller $p$ the mobility edge occurs at larger $V$, while for larger $p$ it moves to smaller $V$.
In Sec.~\ref{S4} we show how this shift correlates with changes in the mode diagram of the Otto cycle.

\section{Quantum Otto cycle}\label{S3}

\begin{figure}[htp]
\includegraphics[width=0.5\textwidth]{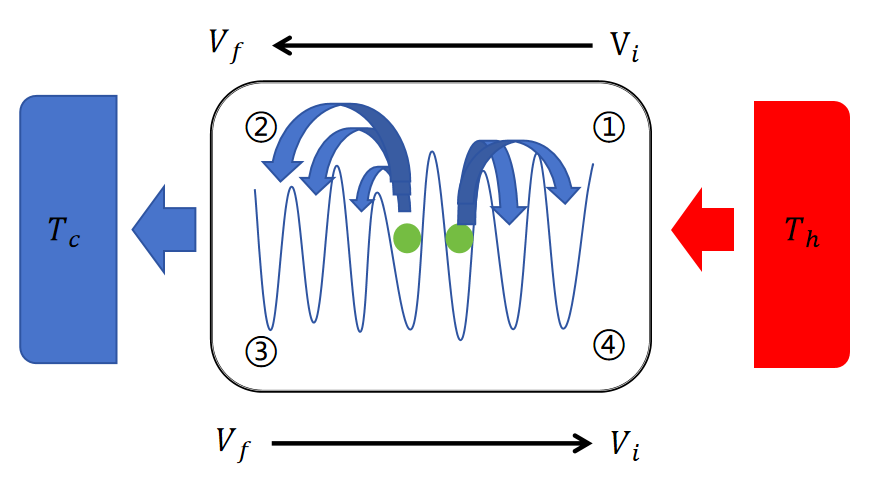}\\
\caption{(Color online) Schematic of the quantum Otto cycle. Strokes $\textcircled{4}\rightarrow\textcircled{1}$ and $\textcircled{2}\rightarrow\textcircled{3}$ are thermalization at fixed Hamiltonians $H(V_i)$ and $H(V_f)$, with reservoirs at $T_h$ and $T_c$.
Strokes $\textcircled{1}\rightarrow\textcircled{2}$ and $\textcircled{3}\rightarrow\textcircled{4}$ are isolated parameter changes from $V_i$ to $V_f$ and back.}
\label{f2}
\end{figure}

We implement a four-stroke Otto cycle using the Biddle--Das Sarma model as the working medium.
A schematic is shown in Fig.~\ref{f2}. The control parameter is the onsite potential strength $V$, which is switched between $V_i$ and $V_f$ during the isolated strokes.
The cycle consists of two thermalization strokes and two isolated strokes, in direct analogy with the classical Otto cycle \cite{ref52,ref55,ref56,ref60}.

During stroke $\textcircled{4}\rightarrow\textcircled{1}$, the system (with Hamiltonian $H(V_i)$) is coupled to a hot bath at temperature $T_h$ and relaxes to thermal equilibrium.
During stroke $\textcircled{2}\rightarrow\textcircled{3}$, the system (with Hamiltonian $H(V_f)$) is coupled to a cold bath at temperature $T_c$ and again equilibrates.
Strokes $\textcircled{1}\rightarrow\textcircled{2}$ and $\textcircled{3}\rightarrow\textcircled{4}$ are carried out under thermal isolation, during which the Hamiltonian parameter is tuned from $V_i$ to $V_f$ and back.
In practice, such isolated strokes need not be perfectly adiabatic because finite-time driving can induce transitions \cite{ref59,ref60}.
Below we consider two limiting protocols that bracket possible behavior.

Let $E_1,E_2,E_3,E_4$ denote the internal energy at the four corners of the cycle.
We define the heat absorbed from the hot bath, the heat released to the cold bath, and the net work output as
\begin{equation}
Q_h = E_1-E_4,\qquad
Q_c = E_3-E_2,\qquad
W = Q_h+Q_c.
\end{equation}
With this convention, $W>0$ corresponds to net work production.
The cycle respects the Clausius inequality by construction.
Following Refs.~[\mycite{ref59,ref60}], we classify the operating mode by the signs of $(Q_h,Q_c,W)$:
\begin{enumerate}
\item \textit{Heat engine}: $Q_h > 0,\; Q_c < 0,\; W > 0$;
\item \textit{Refrigerator}: $Q_h < 0,\; Q_c > 0,\; W < 0$;
\item \textit{Heater}: $Q_h < 0,\; Q_c < 0,\; W < 0$;
\item \textit{Accelerator}: $Q_h > 0,\; Q_c < 0,\; W < 0$.
\end{enumerate}

On the hot isochoric stroke, the density matrix is the Gibbs state
\begin{equation}
\rho_1=\frac{e^{-\beta_h H(V_i)}}{Z_1},\qquad
Z_1=\mathrm{Tr}\!\left[e^{-\beta_h H(V_i)}\right],
\end{equation}
where $\beta_h=(k_B T_h)^{-1}$ and $E_1=\mathrm{Tr}[\rho_1 H(V_i)]$.
On the cold isochoric stroke,
\begin{equation}
\rho_3=\frac{e^{-\beta_c H(V_f)}}{Z_3},\qquad
Z_3=\mathrm{Tr}\!\left[e^{-\beta_c H(V_f)}\right],
\end{equation}
where $\beta_c=(k_B T_c)^{-1}$ and $E_3=\mathrm{Tr}[\rho_3 H(V_f)]$.

We now specify the two protocols for the isolated strokes.

\emph{Near-adiabatic (state-frozen) protocol.}
We assume that the density matrix changes negligibly during each isolated stroke, so the state is approximately ``carried along'' while the Hamiltonian parameter is switched.
This limit can be interpreted as extremely fast switching or as active control that suppresses unwanted evolution, which can be approached using feedback techniques \cite{sayrinRealtimeQuantumFeedback2011,vijayStabilizingRabiOscillations2012,horowitzQuantumEffectsImprove2014,zhangQuantumFeedbackTheory2017,uysQuantumControlMeasurement2018,prechQuantumThermodynamicsContinuous2025}.
We take $\rho_2\simeq \rho_1$ and $\rho_4\simeq\rho_3$, giving
\begin{equation}
E_2=\mathrm{Tr}\!\left[\rho_1 H(V_f)\right],\qquad
E_4=\mathrm{Tr}\!\left[\rho_3 H(V_i)\right].
\end{equation}

\emph{Adiabatic protocol.}
We assume that the isolated strokes are slow enough to preserve level populations in the instantaneous energy basis.
Denote the single-particle energies of $H(V_i)$ and $H(V_f)$ by $\{E_j^i\}$ and $\{E_j^f\}$, respectively.
Because the Hamiltonian is quadratic, equilibrium and mean energies can be evaluated from the single-particle spectrum.
For convenience we use Fermi-Dirac occupations (equivalently, a grand-canonical description with chemical potential absorbed into the energy reference),
\begin{equation}
f_{\beta}(E)=\bigl(1+e^{\beta E}\bigr)^{-1}.
\end{equation}
After thermalization with the hot bath, the occupations are $f_{\beta_h}(E_j^i)$; adiabatic evolution to $V_f$ preserves these occupations while changing the energies, yielding
\begin{equation}
E_2=\sum_{j=1}^{L} E_j^f\, f_{\beta_h}(E_j^i).
\end{equation}
Similarly, after thermalization with the cold bath, the occupations are $f_{\beta_c}(E_j^f)$; adiabatic evolution back to $V_i$ gives
\begin{equation}
E_4=\sum_{j=1}^{L} E_j^i\, f_{\beta_c}(E_j^f).
\end{equation}
These expressions reduce the cycle analysis to the spectra at $V_i$ and $V_f$ and provide a consistent basis for the mode diagrams in Sec.~\ref{S4}.

\section{Multiple thermodynamic modes}\label{S4}

We now present the operating-mode diagrams obtained by evaluating $Q_h$, $Q_c$, and $W$ over parameter space.
We emphasize that the ``mode'' refers only to the sign structure of heat and work, as defined in Sec.~\ref{S3}.
We compare the near-adiabatic (state-frozen) protocol and the adiabatic protocol, and we focus on how the hopping-range parameter $p$ and the tuning amplitude $V_f$ reshape the accessible regions.

\subsection{Near-adiabatic (state-frozen) protocol}

\begin{figure}[htp]
\includegraphics[width=0.5\textwidth]{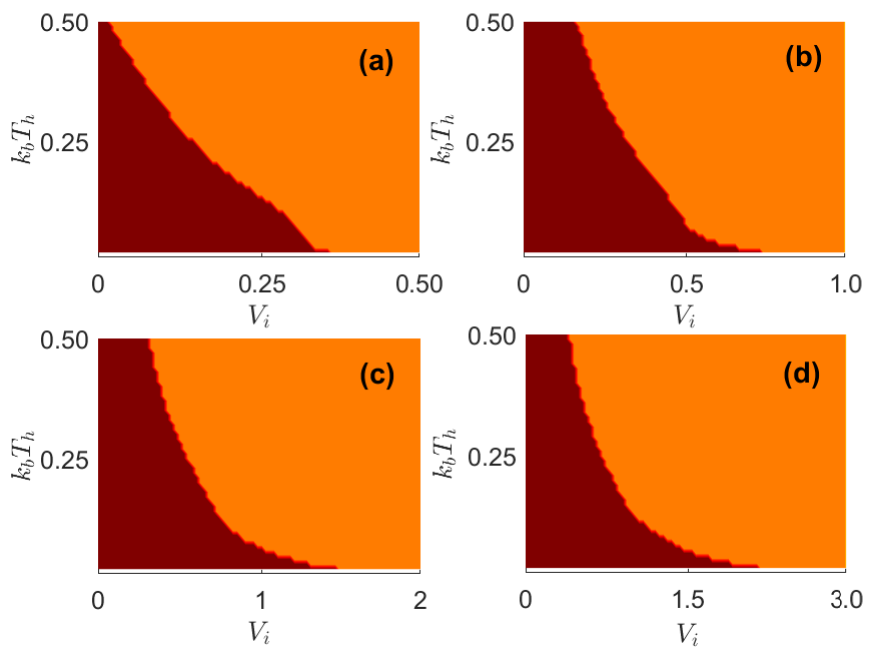}\\
\caption{(Color online) Operating modes in the near-adiabatic (state-frozen) protocol for $p=0.75$ and $L=233$.
(a) $V_f=0.5t$; (b) $V_f=t$; (c) $V_f=2t$; (d) $V_f=3t$.
Brown: \textit{heater}. Yellow: \textit{accelerator}.}
\label{f3}
\end{figure}

Figure~\ref{f3} shows mode diagrams for $p=0.75$ and $L=233$ for several values of $V_f$.
Only two modes appear: \emph{heater} and \emph{accelerator}.
For small $V_f$ [Figs.~\ref{f3}(a) and \ref{f3}(b)], the accessible range of $V_i$ is relatively narrow and the mode boundary is most sensitive to the hot-bath temperature $T_h$.
In this regime, modest changes in $T_h$ can switch the cycle between heater and accelerator operation.
As $V_f$ increases [Figs.~\ref{f3}(c) and \ref{f3}(d)], the boundary shifts toward larger $V_i$, indicating that tuning $V_i$ becomes an effective control knob at fixed $T_h$.
The heater region initially expands with $V_f$ but then approaches saturation: beyond a threshold, increasing $V_f$ mainly enlarges accelerator-dominated regions.
This behavior identifies a practical upper range of $V_f$ for maximizing heater operation without unnecessary parameter excursions.

\begin{figure}[htp]
\includegraphics[width=0.5\textwidth]{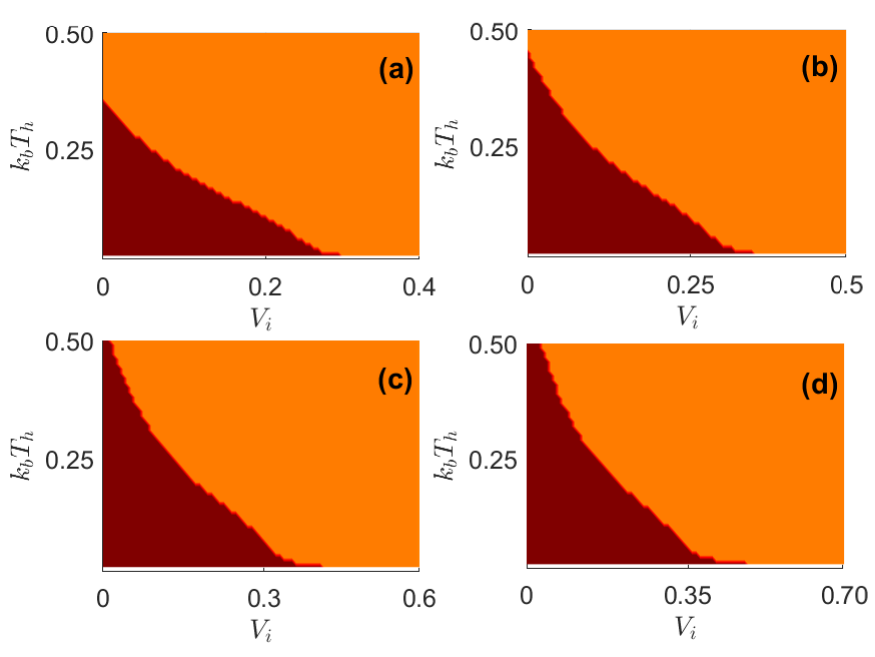}\\
\caption{(Color online) Operating modes in the near-adiabatic (state-frozen) protocol for $p=1.5$ and $L=233$.
(a) $V_f=0.4t$; (b) $V_f=0.5t$; (c) $V_f=0.6t$; (d) $V_f=0.7t$.
Brown: \textit{heater}. Yellow: \textit{accelerator}.}
\label{f4}
\end{figure}

Figure~\ref{f4} shows the corresponding results for $p=1.5$ and $L=233$.
The cycle again supports only heater and accelerator modes, but the heater region is more compressed and the boundary is more tilted in the $(T_h,V_i)$ plane.
This indicates increased sensitivity to \emph{both} $T_h$ and $V_i$: at fixed $V_f$, raising either parameter tends to favor the accelerator mode.
As in the $p=0.75$ case, the heater region grows with $V_f$ and then saturates.
Overall, increasing $p$ reduces the parameter window in which heater operation is possible and lowers the corresponding $T_h$ threshold.
Within this protocol, the mobility-edge structure therefore influences the \emph{extent} and \emph{sensitivity} of the heater region, even though no new modes appear.

\subsection{Adiabatic protocol}

We now turn to the adiabatic protocol, where the isolated strokes preserve level populations.
In this case, the Otto cycle can access additional operating modes.

\begin{figure}[htp]
\includegraphics[width=0.5\textwidth]{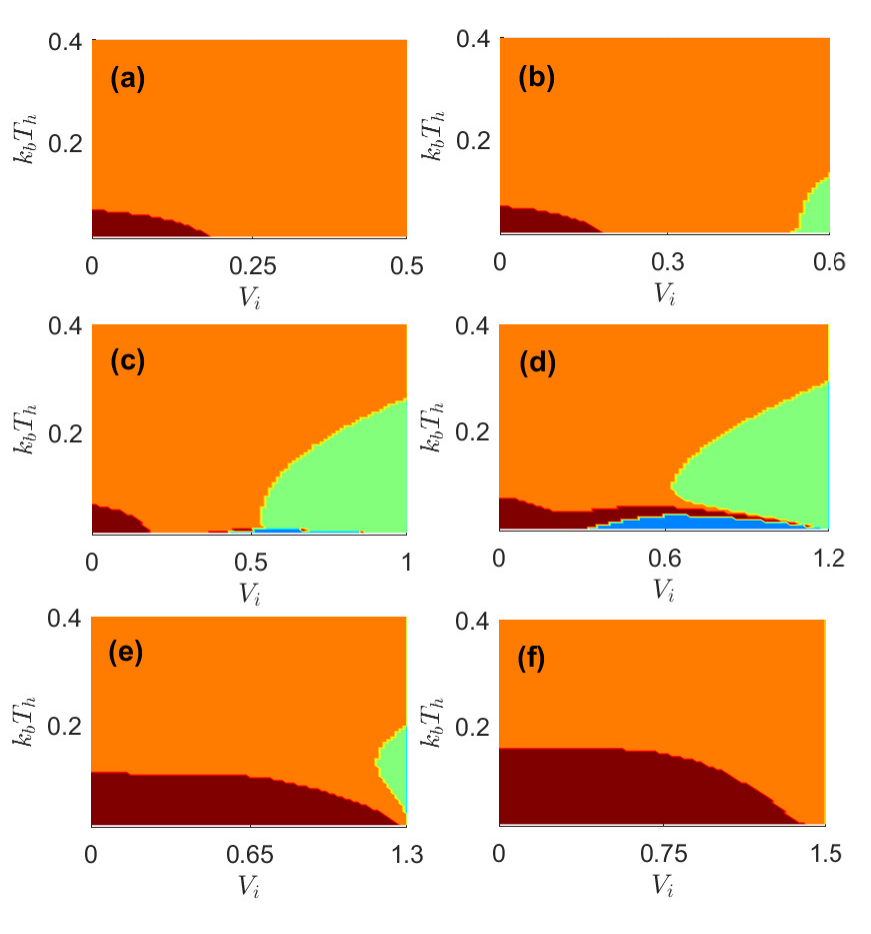}\\
\caption{(Color online) Operating modes in the adiabatic protocol for $p=0.75$ and $L=610$.
(a) $V_f=0.5t$; (b) $V_f=0.6t$; (c) $V_f=t$; (d) $V_f=1.2t$; (e) $V_f=1.3t$; (f) $V_f=1.5t$.
Brown: \textit{heater}. Yellow: \textit{accelerator}. Green: \textit{heat engine}. Blue: \textit{refrigerator}.}
\label{f5}
\end{figure}

Figure~\ref{f5} shows mode diagrams for $p=0.75$ and $L=610$ over a range of $V_f$ values.
For very small and very large $V_f$ [Figs.~\ref{f5}(a) and \ref{f5}(f)], the diagram resembles the near-adiabatic protocol, with only heater and accelerator modes.
At intermediate $V_f$ [Figs.~\ref{f5}(b)--\ref{f5}(d)], two additional modes emerge at low $T_h$: a \emph{heat engine} and a \emph{refrigerator}.
In this regime, mode boundaries become densely packed, and varying $V_i$ at fixed $(T_h,V_f)$ can switch the cycle among three or more modes.
The refrigerator region is the most constrained, requiring both low $T_h$ and a narrow range of $(V_i,V_f)$.
As $V_f$ increases further [Figs.~\ref{f5}(d)--\ref{f5}(f)], the refrigerator region disappears first, while the heat-engine region shrinks more gradually.
The heat-engine region also competes directly with the heater region: when the heat-engine window grows, the heater window is compressed, and vice versa.

\begin{figure}[htp]
\includegraphics[width=0.5\textwidth]{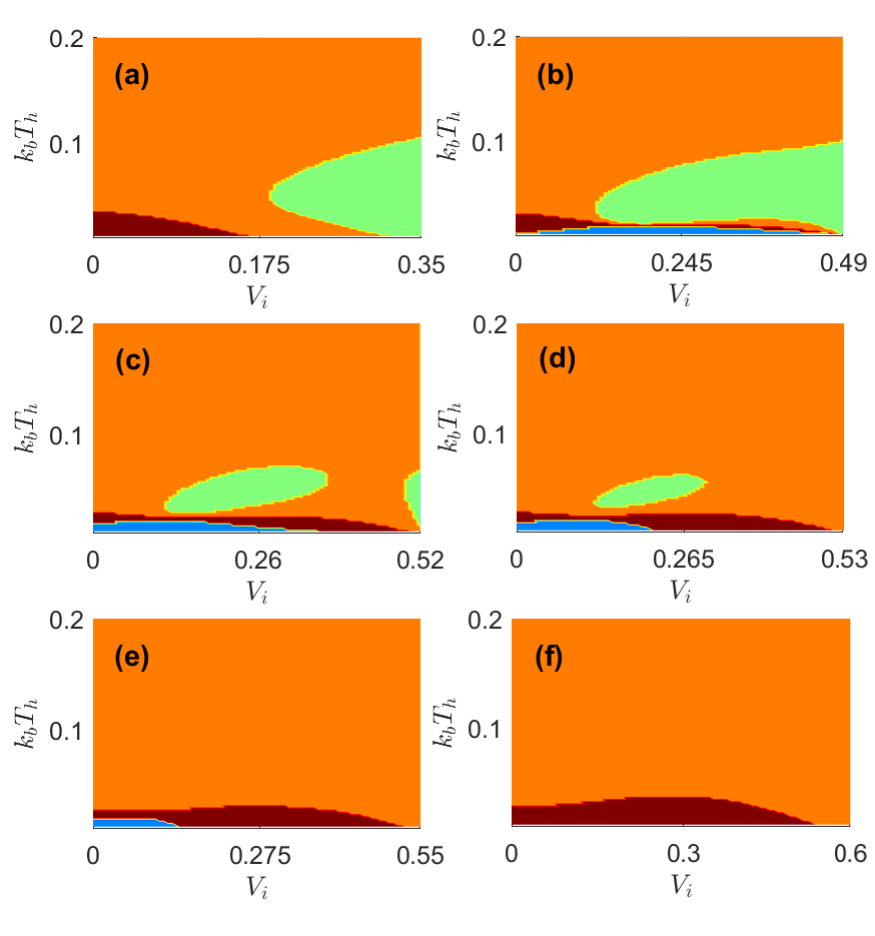}\\
\caption{(Color online) Operating modes in the adiabatic protocol for $p=1.5$ and $L=610$.
(a) $V_f=0.35t$; (b) $V_f=0.49t$; (c) $V_f=0.52t$; (d) $V_f=0.53t$; (e) $V_f=0.55t$; (f) $V_f=0.6t$.
Brown: \textit{heater}. Yellow: \textit{accelerator}. Green: \textit{heat engine}. Blue: \textit{refrigerator}.}
\label{f6}
\end{figure}

Figure~\ref{f6} shows the same analysis for $p=1.5$ and $L=610$.
The high-$T_h$ region is again dominated by accelerator operation, while the low-$T_h$ region supports heater, heat-engine, and refrigerator modes over intermediate $V_f$.
Compared with $p=0.75$, the heat-engine window is more structured: it appears most prominently when $V_i$ and $V_f$ are close and can split as $V_f$ increases.
This splitting is accompanied by a narrowing of the allowable $T_h$ range, indicating stronger parameter constraints for stable heat-engine operation.
In contrast, the refrigerator window can remain visible over a wider span of $V_f$ in the large-$p$ case, suggesting improved robustness of refrigerator operation relative to heat-engine operation in this parameter regime.
As $V_f$ increases, the heat-engine area typically contracts before the refrigerator area does, and the heater region can expand back toward a saturated extent once the competing modes diminish.

Taken together, Figs.~\ref{f5} and \ref{f6} show that the adiabatic protocol unlocks the full set of four thermodynamic modes, but only within
parameter windows that depend on both the spectrum deformation (set by $V_f$) and the hopping range (set by $p$).
In particular, moderate $V_f$ is essential: too small or too large a change tends to suppress the heat-engine and refrigerator windows, returning the cycle to the simpler two-mode behavior.

\section{Summary}\label{S5}

We studied thermodynamic operation of a quasiperiodic mobility-edge lattice, described by the Biddle--Das Sarma model \cite{BS_model}, when used as the working medium of a quantum Otto cycle.
We mapped the operating mode as a function of the hopping-range parameter $p$, the initial and final onsite potentials $V_i$ and $V_f$, and the protocol used for the isolated strokes.

In a near-adiabatic (state-frozen) protocol, the cycle supports only two modes, heater and accelerator, with mode boundaries that shift systematically as $V_f$ is increased and that show clear saturation behavior.
In an adiabatic protocol, the cycle exhibits four modes: heater, accelerator, heat engine, and refrigerator.
The heat-engine and refrigerator windows appear mainly at low $T_h$ and intermediate $V_f$, where competition among modes is strongest.
We further find that increasing $p$ reshapes these windows: the heat-engine region can become more constrained and fragmented, while the refrigerator region can remain comparatively robust in the parameter ranges studied.

Our results highlight that mobility-edge systems can serve as tunable, multifunctional working media for quantum thermal cycles.
An important next step is to broaden the operating windows of the heat-engine and refrigerator modes, for example by refining the cycle protocol beyond the two limiting cases considered here.
Such extensions would help assess how robust these mode diagrams remain under more realistic finite-time driving and experimental constraints.

\begin{acknowledgments}
This research is supported by Zhejiang Provincial Natural Science Foundation of China under Grant No.~LQN25A040012,
the National Natural Science Foundation of China under Grant No.~12174346, and the start-up fund from Xingzhi College,
Zhejiang Normal University.
\end{acknowledgments}

\bibliography{reference}

\end{document}